\documentstyle[sprocl,epsfig]{article}

\def\be{\begin{equation}}
\def\ee{\end{equation}}
\def\bea{\begin{eqnarray}}
\def\eea{\end{eqnarray}}
\begin{document}

\vbox{\baselineskip=14pt%
   \rightline{UH-511-938-99}}

\title{ANALYSES OF LONG LIVED SLEPTON NLSP IN GMSB MODEL AT LINEAR COLLIDER
\footnote{Talk given at the four International Workshop on 
Linear Colliders (LCWS99), April 28 - May 5, Sitges, Barcelona, Spain.}}
\author{P. G. MERCADANTE, H. YAMAMOTO}
\address{
Department of Physics and Astronomy,
University of Hawaii,
Honolulu, HI 96822, USA
}


\maketitle\abstracts{
We performed an analysis on the detection of a long lived stau at a
linear collider with $\sqrt{s}=500$ GeV. In GMSB models a long lived NLSP
is predicted for large value of  the supersymmetry breaking scale $F$.
Furthermore in a large portion of the
parameter space this particle is a stau. Such heavy charged particles
will leave a track in the tracking volume and hit the muon detector.          
 In order to
disentangle this signal from the muon background we explore 
kinematics and particle identification tools: time of flight device,
dE/dX and Cerenkov devices.}

%%%%%%%%%%%%%%%%%% MAIN TEXT%%%%%%%%%%%%%%%%%% 
%%%%%%%%%%%%%%%%%%%%%%%%%%%%%%%%%%%%%%%%%%%%%%%
%\section{Introduction}

In models where a gauge mediated sector is responsible for communicating
Supersymmetry breaking to the MSSM sector~\cite{early,dine} the
gravitino is the lightest supersymmetric particle. Moreover, because
of the very weak interaction of the gravitino, all supersymmetric
particles will decay into the next lightest supersymmetric particle
(NLSP). The life
time of the NLSP can vary from an instantaneous decay to a decay
outside the detector, depending on the value of the Supersymmetry
breaking scale, $F$, which should be considered as a free parameter. 
Moreover, because soft terms in gauge mediated
 models are generated by gauge couplings, the NLSP is either a
neutralino or a stau. 
In this workshop,  the case for a  
neutralino NLSP,
both long and short lived, was considered by Ambrosanio~\cite{work,ambro}, 
and the case for a
stau  NLSP with prompt decay was considered by Kanaya~\cite{work}. 
In this analysis, we studied the search techniques for a stau NLSP in
the production channel $e^+ e^- \rightarrow \tilde{\tau}^+
\tilde{\tau}^- $, in the context of a linear collider at $\sqrt{s}=500$
GeV. The simulation package
ISAJET~\cite{isajet} was used to perform our calculations.

%Within the minimal GMSB model the supersymmetry breaking is
%communicated to the 

%The remaining of
%this  note is organized as follows. In section II we describe the
%ISAJET simulation for the pair production of staus with the effects
%of a time of flight device, dE/dX and a Cerenkov device to detect a
%heavy particle. In section III we use the stau pair production process
%to extract the decay leght of stau. 

%\section{stau Pair Production}

The stau pair production at a linear collider provides a clean search
enviroment.
Futhermore, the production cross section is largely model-independent, 
depending 
only on the mass of the
staus and on the mixing angle between the left and right 
superpartners.       
In Fig.~1a we show the pair production cross section
(normalized to $\sigma_{\mu \mu}=450$ fb) for the left and right
states as a function of stau mass. We can see that
the stau pair cross section is smaller than $\sigma_{\mu \mu}$,
 due to its scalar nature, and
this cross section rapidly drops when we approach the kinematical
limits of the
accelerator. Nevertheless we note that for masses around $240$
GeV we still have cross section of order of $10'$s fb, which  should be
observable provided that the background is manageable.

\begin{figure}[htb]
%\vskip -3cm
\begin{center}
\epsfig{file=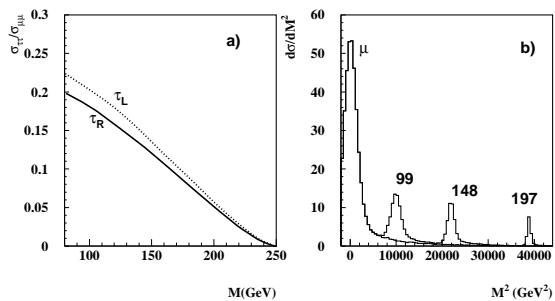,height=8cm}
\end{center}
\caption{a) Stau pair cross section as a function of mass, for a right
and left stau state, normalized by the muon pair cross section. b)
$d\sigma/dM$ for several values of stau masses. Also shown is the $\mu$
background. The width of the peaks reflect the uncertainty in the
momentum measurement.} 
\end{figure}

The signal we are looking for are two back to back tracks with
corresponding hits in the muon chamber. With this requirement
tracks from $\pi, K, p, e$ are removed.  To
reject the two photon process of 
$\gamma \gamma \rightarrow \mu^+ \mu^- $,
we note that the $\mu^+ \mu^-$ pair in this process tends to have a low 
invariant mass
and be boosted along the beam pipe. Thus, we require the following cuts:
\begin{enumerate}
\item{$\cos{\theta} < 0.8$, to guarantee good track quality;}
\item{$|P| > 0.5 E_{\mbox{beam}}$ and} 
\item{$|(P_{\mbox{tot}})_z| < 0.25  E_{\mbox{beam}}$}.
\end{enumerate}
After these cuts, the two photon muon pair
production is estimated to be 1.4 fb. We are then left with muon pair
production $e^+ e^- \rightarrow \mu^+ \mu^-$
as the main source of backgrounds.

In order to reduce the muon pair background we shall explore the heavy mass
of the staus.  In a $e^+ e^-$ collider the energy in the
center of mass is fixed, so in a pair production process the energy of
the final particles is also known. It is well known, however, that
in a high energy linear collider beamstraulung and initial state
radiation  effects become
important and the effective energy of the reaction is not fixed
but presents a spectrum. Nevertheless it is well aproximated by one
photon emission from one of the initial state $e^{\pm}$. With this
aproximation, 
the mass estimate for each track is given by:

\begin{eqnarray}           
M^2&=&\left( \frac{\sqrt{\hat{s}}}{2\gamma} + \beta p_z
 \right)^2 - |p|^2 \;,\\
\hat{s}&=&s(1-|\Delta|) \;,\\
\beta&=&\Delta/(2-|\Delta|) \;,
\end{eqnarray}
where $\Delta=p_z^{\mbox{tot}}/E_{\mbox{beam}}$ is the net momentum in the
beam line direction, $\beta$ is the boost parameter and $\sqrt{\hat{s}}$ is
the center of mass energy of the two tracks.

In Fig.~1b is shown the cross section distribution as a function of
$M^2$ which is estimated according to (1). In this plot we can see the muon
distribution peaking at zero mass with a tail from beamstraulung. We
also see the distribution of staus  production for several values of 
masses. The momentum resolution is taken to be $\delta P_t/P_t=5X10^{-5}
P_t$ (GeV). Based in this plot we use the following cut,

\begin{enumerate}
\setcounter{enumi}{3}
\item {$|M^2-M^2_{\tilde{\tau}}| < 3000 \;\; \mbox{GeV}^2$}; 
\end{enumerate}
Where $M_{\tilde{\tau}}$ is a
variable parameter in the search. The resulting efficiency after this cut
is shown as  the dotted line of Fig~2a.

To futher improve the sensitivity, we study particle identification.
A time of flight device can be used to identify heavy tracks. In our
study we considered a linear collider with $1.4$ ns of bunch
separation. In the large detector  scenario ($r=2$m) the mean time of
flight for a massless ($\beta=1$) particle is around $6.7$ ns. We
assumed that we do not know which bunch crossing a given event is
coming from and 
simulated the effect of a $50$ ps error in the time of flight
measurement. We apply the following cut,
\begin{enumerate}
\setcounter{enumi}{4}
\item {$\Delta t > 0.13$ ns,
where $\Delta t$ is the time of flight difference between a $\beta=1$ and a
massive particle, modulo $1.4$ ns.} 
\end{enumerate}
This cut correspond to about
$2.5 \sigma$, so that about $1\%$ of the muon background  are kept.  
Applying this
cut to both tracks we can relax cuts 2 and 3 which extends the mass
range to the full beam energy. The
efficiency of this cut as a function of mass is shown in Fig.~2a as the
solid line. 
%\begin{figure}[tb]
%\vskip 10cm
%\begin{center}
%\psfig{file=fig2lc.eps,height=8cm}
%\end{center}
%\end{figure}

When a charged particle goes through the detector it deposit
energy by ionization. The amount of energy deposit, dE/dX, is a function of 
$\beta
\gamma$ of the particle~\cite{pdg}.
For a heigh momentum muon going through Argonne we expected  
$dE/dX\sim2.63(\mbox{MeV g}^{-1}\mbox{cm}^{2})$. Based on $5 \%$
resolution  for
$dE/dX$, which is a realistic value for a large TPC, we propose the
following cut, 

\begin{enumerate}
\setcounter{enumi}{5}
{\item  $|dE/dX - 2.63| <
0.4$}; where the units are $(\mbox{MeV g}^{-1} \mbox{cm}^{2})$. 
\end{enumerate}
The resulting efficiency is shown as
the dashed line of Fig.~2a. We note a blind spot for masses around
$150$ GeV.

In Fig.~2b we present, for each strategy, the minimum cross section
that will  be visible
at a $\sqrt{s}=500$ GeV linear collider with $50\;\; {\mbox
{fb}}^{-1}$.  Our criteria is based on a 3 sigma significance; namely,
$S\ge 3$ with
$S=\epsilon \sigma \sqrt{L/bg}$, where $\sigma$ is the
 signal cross
section, $\epsilon$ is the efficiency to pass the cuts and $bg$ is the
expected  background cross section after cuts. We require a minimum of
5 signal events after cuts. 

\begin{figure}[htb]
\begin{center}
\psfig{file=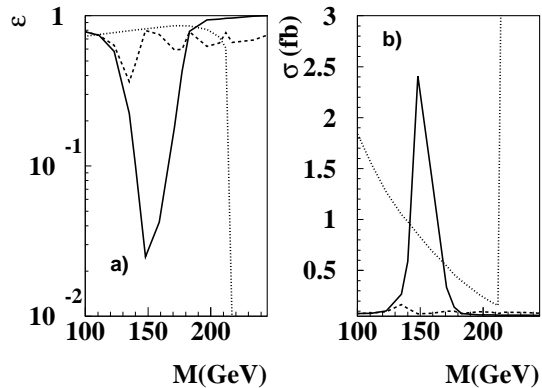,height=8cm}
\end{center}
\caption{a) The efficiency for the signal to pass the various
cuts. The dotted line stands for cuts 1-4, the dashed line for cuts 1
and 5 and the solid lines for cuts 1 and 6. b) Minimum signal cross
section to be observed in a 3 S effect for each set of cuts, same
convention as in a).} 
\end{figure}

A Cerenkov device can be used to measure $\beta \gamma$. With a device
similar to the Babar detector it is
possible to reject particles with $\beta \gamma > 8$. We considered
the case of $4.5 \%$ rejection factor for the light particle (muons)  
while keeping nearly $100 \%$ of the heavy particles. With this
estimative the efficiency of the strategy is the same as using just
the kinematical cuts. The minimum signal cross section, however, 
will be given by the 5 event requirement and the reach will extend all 
the way to the full beam energy.

A  comment on the nature of our results is in order. We have presented a
strategy based only on the pair production mechanism where particle 
identification had a relatively minor role to play. Nevertheless, in the 
models under consideration it is likely that others supersymmetric particles 
will be produced and end up in stau; in such decay chains the use of time of 
flight, dE/dX and Cerenkov devices would play a critical role in identifying 
staus \footnote{The kinematical
distribution, as proposed here, would be of no use in a process other
than pair production.}.

%\section{stau decay length}

Up to this point we supposed a stau that mostly do not decay within
the detector; however, as mentioned
earlier, the life time can be viewed as a free parameter of the
model. We will now discuss a simple measurement of the stau life time
using the mode $e^+ e^- \rightarrow \tilde{\tau^+}  \tilde{\tau^-}$.
The case of a short lived stau is to be studied elsewhere.

\begin{figure}[htb]
%\vskip 10cm
\begin{center}
\psfig{file=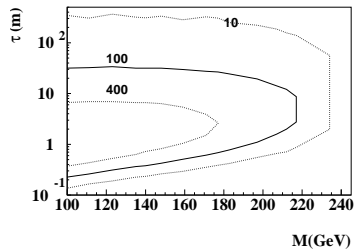,height=5cm}
\end{center}
\caption{Contours of number of tracks decaying between 1m and 2m.} 
\end{figure}

In Fig.~3 we present contours for the number of stau that decay
between radii $1$m and $2$m for the integrated luminosity of $50$
fb$^{-1}$ plotted
as a function of stau mass and life
time.  In order to ensure good measurement of momentum and dE/dX 
 we require that each event should have two tracks longer than 1
meter, using just the mass cut $4$ to select the events. We
believe that such events will be essentially background free: two well
reconstructed back to back tracks, consistent kinematically with a
heavy particle and with at least one of the tracks presenting a kink or fork
from the decay $\tilde{\tau}\rightarrow \tau \tilde{G}$. The figure
indicates that there is a two fold ambiguity for the life time:
for example, for $M=140$ GeV and $100$ decay events in $1 < r < 2$ m
the measured life time is $c\tau=31\pm3$m or $c\tau=0.38\pm0.04$m. 
This ambiguity,
however, is easily resolved by a full fit of the decay distribution.   

In summary, we have studied the long lived stau pair production in a
linear collider at $\sqrt{s}=500$ GeV. Due to the clean kinematics and
the well predicted production cross section, measurement of momentum
alone can detect such particle with masses up to $85 \%$ of the beam
energy. Particle identification devices extend the mass range to
essentially the full beam energy. We have also performed a preliminary
study of lifetime measurement.

%\begin{figure}
%\rule{5cm}{0.2mm}\hfill\rule{5cm}{0.2mm}
%\vskip 2.5cm
%\rule{5cm}{0.2mm}\hfill\rule{5cm}{0.2mm}
%\epsfig{file=fig1lc.eps, height=0.5cm}
%, width=,
%                       bbllx=, bblly=, bburx=, bbury=,
%                       rheight=, rwidth=, clip=, angle=, silent=}%
%\caption{Radiative (off-shell, off-page and out-to-lunch) SUSY Higglets.
%\label{fig:radish}}
%\end{figure}

\section*{Acknowledgments}
This work was supported by the U.S. Department of Energy under grant DE-FG03-94ER40833. P.M. was partially
supported by Funda\c{c}\~ao de Amparo \`a pesquisa do Estado de S\~ao
Paulo (FAPESP).


\section*{References}
\begin{thebibliography}{99}
\bibitem{early} M.~Dine, W.~Fischler and M.~Srednicki, Nucl. Phys. {\bf
B189}, 575 (1981); S.~Dimopoulos and S.~Raby, Nucl. Phys. {\bf B192},
353 (1981); L.~Alvarez-Guam\'e, M. Claudson and M.~Wise,
Nucl. Phys. {\bf B207}, 96 (1982).
%
\bibitem{dine} M.~Dine and A.~Nelson, Phys. Rev. {\bf D48}, 1277 (1993);
 M.~Dine, A.~Nelson, Y.~Shirman, Phys. Rev. {\bf D51}, 1362 (1995);
M.~Dine, A.~Nelson, Y.~Nir and Y.~Shirman, Phys. Rev. {\bf D53}, 2658 (1996).
%
\bibitem{work} See the web page:
http://www.cern.ch/Physics/LCWS99/talks.html 
%
\bibitem{ambro} S.~Ambrosanio and G.A.~Blair hep-ph/9905403
%
\bibitem{isajet} H.~Baer, F. Paige, S.~Protopopescu and X.~Tata,
hep-ph/9810440 (1998).
%
\bibitem{pdg} See, for example, Particle Data Group, Eur. Phys. J. C
{\bf 3},  1 (1998).
\end{thebibliography}
\end{document}